\input{aipcheck}

\documentclass[
    ,final            % use final for the camera ready runs
  ]
  {aipproc}

\layoutstyle{6x9}

\begin{document}

\title{Long Tail of Quantum Decay from Scattering Data}


\author{M. Nowakowski and N. G. Kelkar}{address={Departamento de Fisica, 
Universidad de los Andes, Cra 1E, 18A-10, Bogot\'a, Colombia} }
%\author{N. G. Kelkar}{ address = {Departamento de Fisica, Universidad de los ANdes,A. A. 4976, Bogota, Colombia}} }
\begin{abstract}
Whereas the short time behaviour of an unstable quantum mechanical
system is well understood from its theoretical as well as experimental side,
the long time tail of the very same systems has neither been measured 
experimentally nor is there a theoretical agreement on how to handle it.
We suggest a possible way out of this unsatisfactory state of art.
Theoretically we suggest that the correct spectral function
entering the Fock-Krylov method to calculate the survival amplitude
is proportional to the density of states of a resonance. The latter is
essentially the energy derivative of a phase shift.
As a bonus, we can connect the survival probability to 
scattering data via the phase shift. The method then not only 
establishes the spectral function, but is per se a semi-empirical method
to extract the large time behaviour from scattering data. 
\end{abstract}

\maketitle

%%%%%%%%%%%%%%%%%%%%%%%%%%%%%%%%%%%%%%%%%%%%
%% MAINMATTER
%%%%%%%%%%%%%%%%%%%%%%%%%%%%%%%%%%%%%%%%%%%%

\section{A Short Historical Tale of Quantum Decay}
This year, in 2008, we celebrate the fiftieth anniversary of a famous
quantum mechanical result which has spawned numerous papers on the subject. 
In 1958 Leonid Khalfin proved that in the quantum world, the 
exponential decay law $ P(t)=e^{-\Gamma t}$, 
is only an approximation \cite{khalfin}.
The quantum mechanical survival
probability, without any approximation is, 
\begin{equation} \label{intro1} 
P(t)=\vert A(t) \vert^2=\vert 
\langle \Psi \vert e^{-iHt}\vert \psi \rangle\vert^2 \, .
\end{equation}
It can be calculated at short times to give, 
$P(t) \simeq 1 - (\Delta_{\Psi}H)t^2$, 
which in turn is connected to  
$\frac{dP(t)}{dt}\vert_{t=0}=0 \leftrightarrow
\frac{d(e^{-\Gamma t})}{dt}\vert_{t=0} \neq 0$.
A direct deviation from the exponential decay law at short times has been 
experimentally verified \cite{wilk}.
It is therefore correct to say that we 
have a very good understanding of the first few moments in the life  
of an unstable quantum state.
This is not the case for the large time tail as not only has it never been 
experimentally seen (in spite of efforts to detect it \cite{longtimeexp}), but
its theoretical foundation seems to `enjoy' different treatments 
\cite{urbanow}. 
Our idea presented in this talk is to 
use a semi-empirical method (experimental data in a
theoretical formula) instead of a direct evidence
for the large time behavior of the survival probability.
We will show that this is also closely related to pinning down
a more exact and reliable theoretical framework.  
To achieve the goals we need two different time concepts in Quantum
Mechanics:
(i) Time as a parameter and 
(ii) Time as an observable. 
The first is clearly the variable time $t$ which appears in the Schr\"odinger
equation and the survival probability; the second has to do with 
quantum mechanical observables 
\footnote{Due to Pauli theorem there is no time operator
as a conjugate variable to energy, but time-delay operators can 
exist nonetheless.}
such as delay-time, dwell-time or sojourn-time, 
Larmor-time, traversal-time etc. which were constructed to answer 
questions about the time spent in a region or the quantum collision time. 
In calculating the parametric time dependence, the 
'observable time' will be necessary.
\section{Large Time Behaviour of the Survival Amplitude}
\subsection{The Fock-Krylov Method}
The Fock-Krylov method \cite{FK} is a suitable theoretical framework to 
study the 
large time behaviour of unstable systems. It relies on basic
quantum mechanical results and therefore is, up to a point
which we will discuss later, model-independent.
We first observe that an unstable state $\vert \Psi \rangle$
cannot be an eigenstate to the energy i.e.
$H\vert \Psi \rangle \neq E\vert \Psi \rangle$.
Otherwise the survival amplitude $A(t)$ and the survival probability $P(t)$
would come out trivially to be, 
$A(t)=\langle \Psi \vert e^{-iHt}\vert \Psi \rangle=e^{-iEt}$ and
$P(t)=1$. 
%A consequence: an excited atom cannot be sharp to energy as it
%de-excites i.e. decays.\\
Hence, assuming a continuum,  
$H\vert E \rangle =E \vert E \rangle$, $\langle E'\vert 
E\rangle=\delta(E'-E)$, we are entitled to expand
\begin{equation} \label{FK1}
\vert \Psi \rangle = \int_{\rm Spect(H)}dE \,a(E)\, \vert E\rangle
\end{equation}
where,  
\begin{equation} \label{FK2}
\rho(E)\equiv \frac{{\rm Prob}_{\Psi}(E)}{dE} =
\vert \langle E \vert \Psi \rangle \vert^2=\vert a(E) \vert^2 \, , 
\end{equation}
is a probability density (and as such positive-definite) to find
the states with energy $E$ in the resonance. This distribution is also known as the spectral function. We can now calculate the survival amplitude to obtain
\begin{equation} \label{FK3}
A(t)=\int_{{\rm Spect(H)}}dE \rho(E) e^{-iEt}=\int_{E_{th}}^{\infty}
\rho(E) e^{-iEt}
\end{equation}
which turns out to be a Fourier transform of the spectral function.
$E_{\rm th}$ is the sum of the masses of the decay products.
The success of this general method hinges on the
right choice of $\rho$. There is no general agreement in the
literature on what this function should be (in the next section we
make a claim about the correct choice of $\rho$), but a general 
parameterization looks like, 
$\rho(E)=({\rm Threshold}) \times ({\rm Pole})\times({\rm Form-factor})$, 
i.e., 
\begin{equation} \label{FK4}
\rho(E)=(E-E_{th})^{\gamma}\times P(E)\times F(E) \,.
\end{equation}
Some comments about $\rho$ are in order:
(i) $P(E)$ has a simple pole at $z_R=E_R-i\Gamma_R/2$ 
which leads to the exponential decay law. More poles 
in the fourth quadrant of the complex E-plane 
would modify even the exponential part of the decay.
(ii) $F(E)$ has no threshold and no pole. It is a smooth function which should 
go to zero for large $E$. 
(iii) Large times $t$ correspond in the Fourier transform to
small $E$ (in agreement with the time-energy uncertainty relation). 
Hence, the large time behaviour is due to the
choice of $\gamma$ which is often controversial.
(iv) The transition region between exponential and
non-exponential is partly due to the choice of the form-factor $F$.
(v) The choice of the spectral function is not unique in the literature:
e.g. often $\gamma=0$ and  $f(E)=1$.  
The problem regarding the right choice of the
spectral function will be discussed in the next section. But even
without the explicit knowledge of $\rho$, one can
extract valuable information from the Fock-Krylov method.
We choose the closed path in the complex $E$-plane: 
$C_{\rm R}= C_{\Im}+C_{\Re}+C_{\rm R}^{1/4}$, starting from
zero along the real axis ($C_{\Re}$) attaching to it a quarter of a circle with
radius ${\rm R}$ ($C_{\rm R}^{1/4}$) in the clockwise direction and completing
the path by going upward the
imaginary axis up to zero ($C_{\Im}$). Hence, 
\begin{equation} \label{FK5}
e^{-iE_{th}t}\,I \,\equiv \,e^{-iE_{th}t}\left(\int_{C_{\Re}}...+
\int_{C_{\rm R}^{1/4}}...+\int_{C_{\Im}}...\right) \, ,
\end{equation}
where the dots indicate the integrand from equation (\ref{FK3}) with 
an argument shifted by $E_{\rm th}$
since we start from $0$. The integral we wish to
calculate is along the real axis. $I$ is calculated by using the 
residue theorem with the pole at $z_R$ leading 
to the exponential decay law. We assume that for $R \to \infty$ the 
integral along the arc goes to zero. Thus, 
\begin{eqnarray} \label{FK6}
A(t)&=&A_{E}(t)+A_{LT}(t) \nonumber \\
A_E(t)&=&2\pi i \tilde{P}(z_R) F(z_R)(z_R-E_{th})^{\gamma}e^{iE_Rt}
e^{-\Gamma_Rt/2}\,=\,a_E(t)e^{-\Gamma_Rt/2}
\end{eqnarray}
with $\tilde{P}(z)=\lim_{z \to z_R}P(z)(z-z_R)$ and
\begin{eqnarray} \label{FK7}
A_{LT}(t)&=&({\rm phase})\times \int_0^{\infty}
dx P(-ix+E_{th})F(-ix+E_{th})x^{\gamma}e^{-xt}\nonumber \\
&\simeq &({\rm phase})\times \Gamma(\gamma +1)P(E_{Th})F(E_{th})
\times  \frac{1}{t^{\gamma +1}} 
=a_{LT}\frac{1}{t^{\gamma +1}}
\end{eqnarray}
The result agrees with \cite{Fonda}. This is how nature slows
down the exponential decay. 
\subsection{The transition region and critical times}
One can approximately estimate the transition time from the 
exponential to the power law behaviour by setting
$\vert a_E \vert e^{-2\Gamma_Rt_0/2} \simeq \vert a_{LT}\vert 
\frac{1}{t_0^{\gamma +1}}$
or alternatively by determining the zeros of the function
\begin{equation} \label{FK8}
\omega(\xi_0)\equiv \ln
\frac{\vert a_E\vert}{\vert a_{LT}\vert}
\left(\frac{\Gamma_R}{2}\right)^{-\gamma-1}
+(\gamma+1)\ln\ \xi_0 -\xi_0, \,\,\, \xi_0 \equiv \frac{\Gamma_Rt}{2}
\end{equation}
\begin{figure}
\includegraphics[height=0.2\textheight]{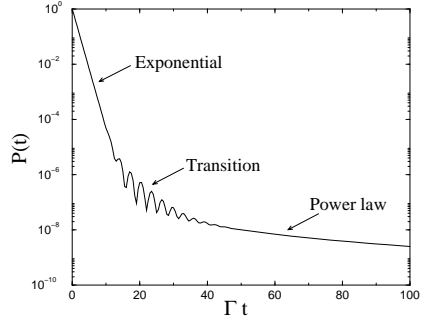}
\caption{Characteristic behaviour of a survival probability.}
\end{figure}
Note that, strictly speaking, there are three regions (see Fig. 1) and two
critical times: the time at the transition from the exponential to the 
oscillatory region and from the oscillatory to the power law. The condition
(\ref{FK8}) can also have two zeros out of which normally the second one 
is the right indicator of critical time. 
Consider a narrow resonance model with the spectral function 
a simple Breit-Wigner,
$\rho(E) \propto (E_R - E_{th})^{\gamma}
\frac{e^{-E/E_0}}{(E-E_R)^2+(\Gamma_R/2)^2}$. 
Then defining 
$\epsilon \equiv E_R-E_{th}$ 
and choosing $\gamma=1/2$
\begin{equation} \label{FK9}
\frac{\vert a_{E}\vert}{\vert a_{LT}\vert (\Gamma_R/2)^{3/2}}
\propto e^{-\epsilon/E_0}\left[1 + \epsilon^2/(\Gamma_R/2)^2\right]^{5/4}\, .
\end{equation}
It is evident that smaller the ratio $\epsilon/\Gamma_R$, smaller is 
the critical time\footnote{This is an approximate estimate since there are
two critical times characterizing the transition region}.
Hence for narrow resonances the best candidates
to find the non-exponential long time behaviour are
threshold resonances.
%\begin{figure}
  %\includegraphics[height=.3\textheight]{golfer}
  %\caption{Picture to fixed height}
%\end{figure}
%\section{The Connection to Scattering Data: Beth-Uhlenbeck Density of States}
\section{The Connection to Scattering Data}
Beth and Uhlenbeck \cite{BU}, while calculating
virial coefficients $B, C$ 
in the equation of an ideal gas: $pV=RT\left[1 +\frac{B}{V} + 
\frac{C}{V^2} + \cdots \right]$, found 
that the difference between the density of states 
(of scattered particles) with interaction $dn_l(E)/dE$ 
and without $dn^{(0)}_l(E)/dE$ is,  
\begin{equation} \label{BU1}
\frac{dn}{dE}=\frac{dn_l(E)}{dE} -\frac{dn^{(0)}_l(E)}{dE} =\frac{2l+1}{\pi}
\frac{d\delta_l(E)}{dE}\, .
\end{equation}
In a resonant scattering \cite{we2} this is the density of states 
of a resonance (in terms of the decay products). For instance
$T=(\Gamma_R/2)/(E_R -E -i\Gamma_R/2)$, gives, 
\begin{equation} \label{BU2} 
\frac{d\delta}{dE}=\frac{\Gamma_R/2}{(E_R-E)^2 +\Gamma_R^2/4} \, .
\end{equation}
This offers a semi-empirical method to examine the 
large time behaviour of 
of unstable systems (resonances) directly from data 
since the spectral function is, 
\begin{equation} \label{BU3}
\rho(E)=\frac{d{\rm Prob}}{dE}=\frac{1}{n}\frac{dn}{dE} \, .
\end{equation}
Note that this also uniquely fixes the spectral function, 
at least in the vicinity of a resonance.
This interpretation works well for all $l$- values except for the  
$s$-wave, because in this case
$\frac{d\delta}{dE} \propto \frac{1}{(E-E_{th})^{1/2}}$ 
and we encounter a threshold singularity.
This unreasonable singular behaviour of the density of states can be 
remedied without changing
the interpretation (see next section). Otherwise for $l > 0$ we have 
$\frac{d\delta}{dE} \propto (E-E_{th})^{l-1/2}$ and therefore
$\gamma = l-1/2$, i.e., $\gamma$ is fixed.
An explicit example of $P(t)$ for narrow resonances $\alpha + \alpha
\to \,  ^8Be(l=2) \to \alpha + \alpha$
has been calculated in \cite{we} from scattering data given 
in form of the phase shift. 
\section{Explicit Examples}
The Fock-Krylov method is not limited to narrow resonances.
We can put forth the question if there exist new features in the survival
probability for broad resonances? 
For narrow resonances we can use a Breit-Wigner
model for the amplitude to calculate the critical time and obtain 
some relevant results.
No general parameterization of the transition amplitude for broad 
resonances exists.
We will therefore choose the most prominent example of the $\sigma$ for which 
some parameterizations are available. 
The new features which can appear here are e.g., 
sub-threshold zeros (called Adler's zeros) 
in the amplitude and hence also in the form-factor 
in the density of states. Besides this, the $\sigma$ is an s-wave resonance.
As we already mentioned the s-wave density of states has to be modified from 
the time delay to the dwell time delay.
\begin{equation} \label{E1}
\left(\frac{dn}{dE}\right)_{\rm new}={\rm dwell-time}=
2\frac{d\delta}{dE} -\frac{2\Re e (T)\sqrt{s}}{s-4m_{\pi}^2} 
\end{equation}
which is the relativistic version of the expression found in \cite{dwell}. 
This singularity-free expression is also a density of states as shown in
\cite{newden}.
There exist different parameterizations of the 
amplitude for $\pi \pi \to \sigma \to \pi \pi$. Here we have opted 
for the following one from \cite{Bugg}: 
\begin{equation} \label{E2}
T=\frac{M\Gamma(s)}{M^2 -s -iM\Gamma(s)} \, ,  
\end{equation}
where the energy dependent width can be found in \cite{Bugg}.
It has the structure: threshold $\times$ Adler's zeros $\times$ form-factor. 
\begin{figure}
\includegraphics[height=0.35\textheight]{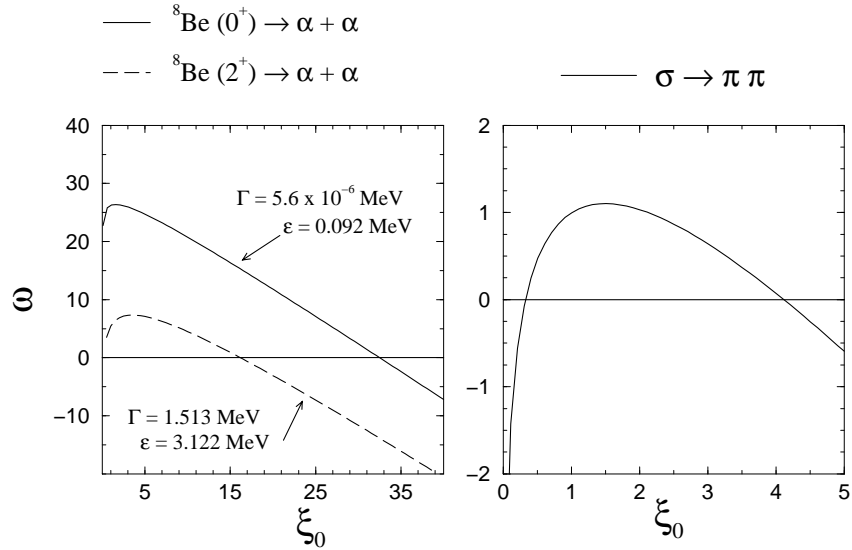}
\caption{Critical times for narrow resonances and $\sigma$ as an example
of a broad resonance.}
\end{figure}
Figure 1 displays the critical times ($\omega(\xi_0)$
from equation (\ref{FK8})) for very narrow resonances in nuclear 
physics ($^8Be$) in contrast to the same exercise done 
for the broad $\sigma$.
The critical lifetimes of these narrow resonances
are indeed large ($\sim 30$ and $70$ in terms of lifetimes). 
However, in the case of a broad resonance, i.e 
$\sigma$, we can conclude the following.
(i) The short transition time which is usually considered an artifact
of the approximation (recall that in reality there are two critical times)
is the biggest we find (ca. one lifetime). The large transition time 
which usually is the correct time scale for transition is the smallest we 
find (eight lifetimes).
Therefore, we think that
the oscillatory transition region could here be of importance and 
the real transition time could be somewhere between the two we find now.
This behaviour would be similar in other broad resonance systems
such as the $\eta$-mesic nuclei \cite{eta}.
The relatively small critical time and the importance of the oscillatory 
region makes the study of the time evolution of the $\sigma$ an interesting 
undertaking which we plan to continue in future.

%\endinput
\end{document}